\documentclass{PoS}

\title{Applications of baryon chiral perturbation theory.\\ 
A topical example: The nucleon sigma terms}

\ShortTitle{Applications of B$\chi$PT: The nucleon $\sigma$-terms}

\author{\speaker{J. Martin Camalich}\\
        Department of Physics and Astronomy, University of Sussex, BN1 9QH, Brighton, UK\\
        E-mail: \email{J.Camalich@sussex.ac.uk}}

\abstract{We present an overview of modern approaches to low-energy baryon structure based on baryon chiral perturbation theory. These are driven by the emergence of Lorentz covariant schemes and the systematic consideration of the effects of the lowest-lying decuplet resonances. In order to illustrate the progress recently achieved in this field, we present the last developments on our understanding of the nucleon sigma terms along these lines. In particular, we will show how these methods, in $SU(2)$ or $SU(3)$ settings, are reliable tools to process and maximize the information on the physical structure of the nucleon one can obtain from either experimental data or lattice QCD results.}

\FullConference{The 7th International Workshop on Chiral Dynamics,\\
		August 6 -10, 2012\\
		Jefferson Lab, Newport News, Virginia, USA}

\begin{document}

One of the major scientific challenges in fundamental physics consists of understanding the strong interactions and the nuclear structure phenomena, directly from QCD. The low-energy structure of the nucleons and other baryons, and their interaction with the Goldstone bosons of chiral symmetry (pions, kaons and eta's), turns out to be an essential ingredient of this program. This is, in fact, a topic much revitalized in recent years, mainly due to the remarkable progress made by the lattice QCD (LQCD) community on the calculation of key observables~\cite{WalkerLoudCD}. This is being facilitated, at the same time, by the emergence of modern effective field theory approaches, based on baryon chiral perturbation theory (B$\chi$PT), which allow to clear the LQCD simulations from some systematics and check them against the experimental data, or to process efficiently the information generated to provide reliable predictions. On the other hand, investigations involving accurate determinations of baryonic observables (such as the proton electric radius or the sigma term) have become part of the modern ``new-physics rush'', and interest in this direction is rapidly spreading across the field.

In this contribution I will present a snapshot of the improvements in B$\chi$PT by reporting the progress recently made on the understanding of the nucleon sigma terms and $\pi N$ scattering. These observables can be studied using either $SU(2)$ or $SU(3)$ settings and can be extracted from experimental data on $\pi N$ scattering and the baryon-octet mass splittings or LQCD results on the lowest-lying baryon spectrum. Therefore, they represent a perfect example to study the potential and limitations of B$\chi$PT for processing and correlating all this information in a model-independent and systematic manner. Moreover, the sigma terms epitomize the type of baryonic observables with high physics interest, containing information on the origin of the mass of the ordinary matter as well as becoming the main uncertainty in the constraints derived from direct searches of dark matter~\cite{Bottinoetal}. 

One customarily introduces two independent observables, $\sigma_{\pi N}$ and  $\sigma_s$, which are known as the pion-nucleon and nucleon strangeness sigma terms. These are defined in the isospin limit ($m_u=m_d\equiv\hat{m}$) as
\begin{eqnarray}
&&\sigma_{\pi N}=\langle N|\hat{m}\left(\bar{u}u+\bar{d}d\right)|N\rangle,\nonumber\\
&&\sigma_s=\langle N|m_s\bar{s}s|N\rangle.\label{Eq:STsDefinition}
\end{eqnarray}
and contain information on the contribution of the quark-condensate to the masses of the baryons and parametrize the flavour-structure of the nucleon scalar form factors at $t=0$. The $\sigma_s$ is of particular significance as it contains information on the virtual $s\bar{s}$ pairs and their contribution to the nucleon mass. In the following we will briefly review the state-of-the-art in B$\chi$PT and, then, report recent determinations of these matrix elements using this approach in combination with experimental data or LQCD results.

\section{Power Counting and decuplet resonances in B$\chi$PT}

Unlike in the meson sector, in B$\chi$PT the power counting (PC) is violated by the presence of $M_N$ as a heavy scale and the baryonic loop diagrams do not fulfill the chiral order prescribed by their topology~\cite{Gasser:1987rb}. A crucial observation follows from noticing that this naive PC arises from considering the nucleons in a non-relativistic expansion, which eventually can be implemented from the outset using heavy-baryon (HB)$\chi$PT~\cite{Jenkins:1990jv}. The HB is an elegant formalism with a neat PC, but it alters the analytic structure of the baryon propagators. This has been argued to be the reason behind the problematic convergence showed by the HB expansion in some observables, motivating the study and emergence of Lorentz covariant approaches. An important development in this sense comes from realizing that the genuine non-analytical chiral corrections in a covariant formalism satisfy the PC and are identical to those obtained in the HB expansion. The PC-breaking pieces, on the other hand, are analytic and can be eventually absorbed into the local counterterms or low energy constants (LECs) in some (renormalization) prescription~\cite{becher,scherer1}. There are two main covariant approaches, the infrared (IR)~\cite{becher} and the extended-on-mass-shell (EOMS) scheme~\cite{scherer1}. The former one, however, introduces unphysical cuts that can disrupt the chiral expansion. The EOMS scheme, on the other hand, is nothing else than conventional dimensional regularization in which the finite parts of the counter-terms are adjusted to cancel the PC-breaking pieces. In this way, it recovers the PC at the same time as it does not change the analytic structure dictated by $S$-matrix theory. 

Another difficulty in B$\chi$PT is the treatment of the lowest-lying decuplet resonances. In the conventional approach, these resonances and other heavier degrees of freedom are integrated out and accounted for by the LECs. This is a valid prescription as long as the energies probed in the theory are well below the mass gap of these states with respect to the ground state octet baryons, or $M_N$. However, in case of the decuplet resonances like the $\Delta(1232)$, the mass gap is only of $\delta\sim M_\Delta -M_N\sim 300$~MeV and, moreover, this resonance couples strongly to the $\pi N$ system. In a $SU(3)$-B$\chi$PT approach, the size of the perturbative parameter, $\sim M_K/\Lambda_{\chi SB}$ is even larger than this typical scale $\delta/\Lambda_{\chi SB}$. Therefore, it becomes necessary to properly take the $\Delta(1232)$ and other decuplet resonances into account as explicit degrees of freedom. In order to do so, one needs to define a suitable PC for the new scale $\delta$~\cite{Hemmert:1997ye,Pascalutsa:2002pi}, and also to tackle the {\it consistency} problem that afflicts interacting spin-3/2 theories (see Ref.~\cite{Pascalutsa:2006up} and references therein). Once these two issues are solved, one can apply any of the formalisms to cure the power counting problem and explicitly calculate their contributions to low-energy baryon structure. 

\section{Experimental determinations of the sigma terms}  
\subsection{The $\pi N$ scattering amplitude and $\sigma_{\pi N}$}   

The elastic scattering of pions and nucleons probes their scalar vertices, and this is formalized in the nucleon case by the Cheng-Dashen theorem connecting  $\sigma_{\pi N}$ to the isospin-even scalar scattering amplitude at the kinematical point $(s=m_N^2,\,t=2M_\pi^2)$, which lies in the unphysical region of the process. The traditional extrapolation is done using an energy-dependent parameterization of the data in partial waves (PW) supplemented by dispersion relations that impose strong analyticity and unitarity constraints onto the scattering amplitude at low energies. There are two ``classic'' determinations of $\sigma_{\pi N}$, the one based on the Karlsruhe-Helsinki (KH) group, $\sigma_{\pi N}\simeq 45(8)$~MeV~\cite{KA85,sigmatermupdate}, and the other performed by the George-Washington (GW) group, $\sigma_{\pi N}=64(7)$~MeV~\cite{WI08,Pavan:2001wz}. Although the difference between these two determinations is not too large, it leads to radically different interpretations of the strangeness content in the nucleon, as we will see below. Besides, a substantial reduction of the $\sim30$ MeV uncertainty involved by these two determinations would increase the significance of the constraints set on the parameter space of models from the experimental bounds on the dark-matter nucleon cross sections.    

In order to understand some of the systematic effects, one would wish to complement the dispersive treatments with B$\chi$PT. Ideally, one would even dream of performing a completely model-independent analysis of the scattering data 
leading to a systematic study of the subthreshold region, $\sigma_{\pi N}$ and all other related quantities without any further input. However these studies have faced important difficulties. The classical works of Fettes et {\it  al.} in HB at $\mathcal{O}(p^3)$~\cite{fettes3} and $\mathcal{O}(p^4)$~\cite{fettes4} were able to reproduce the $S$- and $P$-wave phase shifts in the threshold region, but they didn't succeed to give a realistic description of the subthreshold region and, consequently, they overestimated the value of the sigma-term. They concluded that an order-by-order improvement in the extrapolation onto the subthreshold region was far from obvious~\cite{fettes4}. The inclusion of the $\Delta$ as explicit degree of freedom in the small-scale-expansion (SSE) ($\delta\sim \mathcal{O}(p)$) allowed to stretch significantly the reproduction of the phase-shifts to larger energies~\cite{fettesD}. However, large correlations among the LECs were reported, with values depending much on the PW analysis used as input. As a result a stable value of $\sigma_{\pi N}$ and extrapolation to the subthreshold region could be only achieved using the Olsson dispersive sum rules~\cite{Olsson}. These problems were corroborated by the $\mathcal{O}(p^4)$ calculation without explicit $\Delta$'s done in the IR scheme~\cite{beche2}. In this case, though, an inverse approach was followed. The subthreshold description was investigated and the extension into the physical region was then attempted, without success. 

The situation has recently improved with a novel chiral analysis of the $\pi N$ scattering amplitude introducing two main innovations over previous work. In the first place, a fully covariant approach in the EOMS scheme is employed~\cite{Alarcon:2011zs,Alarcon:2012kn,Chen:2012nx}. This proves to be not only important in the extrapolation onto the subthreshold region (in comparison with HB) but also in extending the framework to higher energies (as the IR scheme becomes sensitive to its unphysical cuts). The second essential ingredient is the inclusion of the $\Delta(1232)$ as an explicit degree of freedom in the $\delta$-counting~\cite{Pascalutsa:2002pi}, which exploits the hierarchy $M_\pi<\delta<\Lambda_{\chi SB}$ by counting $\delta$ as $\mathcal{O}(p^{1/2})$. This analysis was performed up to $\mathcal{O}(p^3)$ in this counting, implying that the only explicit $\Delta$ contributions are those stemming from the Born-Term diagrams. The strategy followed in this work was to determine the different LECs with the $S$- and $P$-wave phase shifts provided by the KH, GW and Matsinos'~\cite{EM02} groups and then discuss the resulting phenomenology.

\begin{figure}[t]
\centering
\includegraphics[width=10cm]{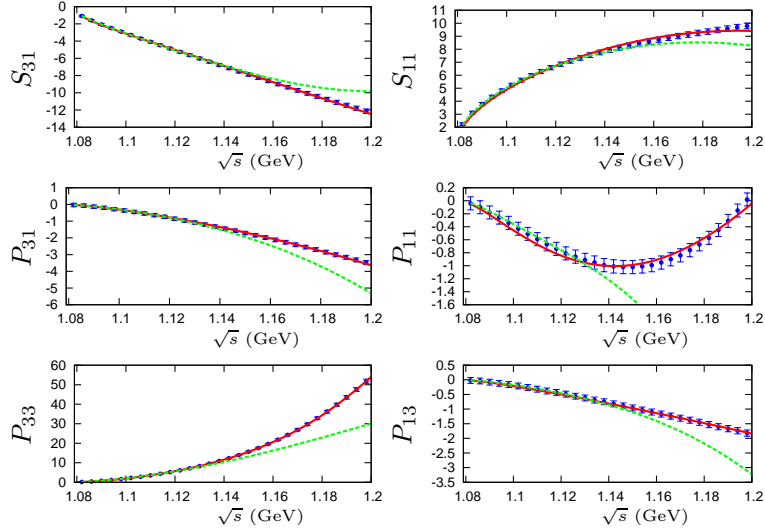}
\caption{Chiral analysis of the GW phase shifts (blue points) up to $\mathcal{O}(p^3)$ in the EOMS scheme without $\Delta$ (dashed green line) and with $\Delta$ (solid red line) in the $\delta$-counting.\label{Fig:piNPSs}}
\end{figure}

The quality of the corresponding fits to the GW PW analysis is shown in Fig.~\ref{Fig:piNPSs} where the phase shifts are perfectly reproduced up to energies of $W=\sqrt{s}\simeq1.2$ GeV. For the sake of comparison, we also include the result obtained without the explicit $\Delta$ contribution, which achieves a good description of the phase shifts only up to energies slightly above threshold.\footnote{The inclusion of the $\Delta$ explicitly up to $\mathcal{O}(p^3)$ in the $\delta$-counting introduces 3 new LECs through the Born-term diagram. However, one of these parameters can be fixed with the $\Delta(1232)$-resonance width and the other two can be shown to be redundant~\cite{Long:2010kt,Alarcon:2012kn}.} In fact, a comparison between the contributions at different orders shows that only in the former case a good convergence is obtained up to $\mathcal{O}(p^3)$ in all the low-energy region. 

\begin{table}
\centering
\caption{Physical observables obtained from the $\mathcal{O}(p^3)$ $\pi N$ scattering amplitude in the EOMS renormalization scheme fitted to different PW analyses. We show the $\pi N \Delta$ coupling $h_A$, the $\pi N$ coupling and corresponding Goldberger-Treiman discrepancy $\Delta_{GT}$, the scattering lengths, which are given in units of 10$^{-2}$ $M_\pi^{-1}$ and the pion-nucleon sigma term, which is given in units of MeV~\label{Table:Observables}}
\begin{tabular}{|cccccccc|}
\hline
&$\chi^2_{\rm d.o.f.}$&$h_A$&$g_{\pi N}$&$\Delta_{GT}$ [\%] &$a_{0+}^+$ &$a_{0+}^-$&$\sigma_{\pi N}$ \\
\hline
KH~\cite{KA85}&0.75&3.02(4)&13.51(10)&4.9(8)&$-1.2(8)$&8.7(2)&43(5)\\
GW~\cite{WI08}&0.23&2.87(4)&13.15(10)&2.1(8)&$-0.4(7)$&8.2(2)&59(4)\\
EM~\cite{EM02}&0.11&2.99(2)&13.12(5)&1.9(4)&0.2(3)&7.7(1)&59(2)\\
\hline
\end{tabular}
\end{table}

Once the LECs are determined, one can predict and study different $\pi N$ scattering observables or to investigate the extrapolation onto the subthreshold region. In Table~\ref{Table:Observables} we present the results for some selected observables, that can be checked to be in good agreement with those reported by the respective PW analyses. Then, one can see that a B$\chi$PT analysis of the phase shifts ratifies the discrepancy between different PW analyses, in particular regarding the value of $\sigma_{\pi N}$. A comparison of the values of some of these observables with the alternative determinations that can be obtained from other sources like pionic atoms or $NN$-scattering favors the GW solution. The KH solution gives rise to a value for $h_A$ that is not compatible with the $\Delta(1232)$ width and to a value for $g_{\pi N}$ that leads to a significant violation of the GT relation. As for our study of the EM PW analysis, we found a value for the isovector scattering length that is too small as compared with the accurate values obtained from pion-atoms data~\cite{EM02}. However, the most important observable in the discussion of $\sigma_{\pi N}$ concerns the scalar-isoscalar scattering length. While the KH result is compatible with the old negative values, it is not with the more recent determinations from modern pionic-atom data, $a_{0+}^+=-0.1(1)$ $10^{-2} M_\pi^{-1}$~\cite{Baru:2010xn}.\footnote{See Ref.~\cite{Alarcon:2012kn} for details.} These are, on the other hand, compatible with the scattering data extractions from the GW and EM solutions. Finally, notice that the Matsinos and GW analyses lead to the same $\sigma_{\pi N}$. This is relevant because these are two completely different analyses that incorporate the new high quality data collected over the 2 last decades, whereas the KH group stopped updating theirs in the mid 80's. With these considerations, we reported the value~\cite{Alarcon:2011zs}
\begin{equation}
\sigma_{\pi N}=59(7) {\rm MeV},\label{Eq:SpiNValue}
\end{equation}
where the error includes a theoretical uncertainty estimated with the explicit calculation of higher-order graphs added in quadrature with the one given by the dispersion of the values in the average of the GW and EM results.

\begin{table}%[H]
 \begin{center}
\begin{tabular}{|c||c|c|}
\hline
                                      & B$\chi$PT          & Dispersive              \\   
\hline
 $d_{00}^+$ ($M_\pi^{-1}$)            & $-1.48(15)$ &      $-1.46$                   \\    
 $d_{01}^+$ ($M_\pi^{-3}$)            & $1.21(10)$   &      $1.14$                   \\    
 $d_{10}^+$ ($M_\pi^{-3}$)            & 0.99(14)   &        1.14(2)                  \\                                                        
 $d_{02}^+$ ($M_\pi^{-5}$)            &  0.004(6)   &       $0.036$                  \\                                                        
 $b_{00}^+$ ($M_\pi^{-3}$)            & -5.1(1.7)    &      $-3.54(6)$               \\                                                        
 $d_{00}^-$ ($M_\pi^{-2}$)            &  1.63(9)  &          1.53(2)                 \\                                                        
 $d_{01}^-$ ($M_\pi^{-4}$)            & -0.112(25)  &       $-0.134(5)$              \\                                                        
 $d_{10}^-$ ($M_\pi^{-4}$)            &  -0.18(5)  &        $-0.167(5)$              \\                                                        
 $b_{00}^-$ ($M_\pi^{-2}$)            & 9.63(30)   &         10.36(10)                \\                                                        
\hline		   	   
\end{tabular}
{\caption[pilf]{Results for different subthreshold coefficients obtained from the fits to the KH analysis and in B$\chi$PT in the EOMS scheme and with explicit $\Delta$ contributions up to $\mathcal{O}(p^3)$ in the $\delta$-counting. The results obtained using dispersive techniques are included for the sake of comparison.
\label{subthreshold-results}}}
\end{center}
\end{table}

The success of this modern calculation to provide a reliable determination of $\sigma_{\pi N}$ from phase shifts can be understood by analyzing the scattering amplitude in the subthreshold region, where it comes usually characterized by the so-called subthreshold coefficients stemming from a kinematic expansion about the point $s=u=m_N^2+M_\pi^2$ and $t=0$. In Table~\ref{subthreshold-results} we show the values of these quantities obtained after fitting the LECs to the KH phase shifts compared to those given by dispersive techniques~\cite{Alarcon:2012kn}. As it can be seen, in this approach to B$\chi$PT, the long-sought connection between the physical and subthreshold regions is accomplished. In particular, $d_{00}^+$ and $d_{01}^+$ correspond to the leading orders of the expansion in $t$ of the Born-subtracted scalar-isoscalar amplitude and so they are essential to understand the extrapolation to the Cheng-Dashen point~\cite{sigmatermupdate} and determination of $\sigma_{\pi N}$. For the details on the discrepancy on $d_{02}^+$ and its meaning we refer the reader to Ref.~\cite{Alarcon:2012kn}.

\subsection{The baryon-octet mass splittings and the strangeness content of the nucleon}

The contribution of the strange quark to the nucleon mass can be related with $\sigma_{\pi N}$ and the $SU(3)_F$-breaking of the baryon masses in the octet. Namely, one can re-express the pion-nucleon sigma term as~\cite{Gasser:1980sb}
\begin{equation}
\sigma_{\pi N}=\frac{\sigma_0}{1-y}, \label{Eq:sigmapiN-sigma0-y}
\end{equation}
where $y$ is the so-called ``strangeness content'' of the nucleon,
\begin{equation}
y=\frac{2\langle N|\bar{s}s|N\rangle}{\langle N|\bar{u}u+\bar{d}d|N\rangle}=\frac{2\hat{m}\sigma_{s}}{m_s\sigma_{\pi N}}=1-\frac{\sigma_0}{\sigma_{\pi N}},\label{Eq:Defy}
\end{equation}
and
\begin{equation}
\sigma_0=\hat{m}\langle N|\bar{u}u+\bar{d}d-2\bar{s}s|N\rangle.\label{Eq:Defsigma0}
\end{equation}
Thus, $\sigma_0$ can be understood as the value of the pion-nucleon sigma term in case that the strange-quark contribution to the nucleon wave function is null (Zweig rule). The interest of $\sigma_0$ lies on the fact that it can be calculated in $SU(3)_F$ B$\chi$PT up to $\mathcal{O}(p^3)$ using the experimental baryon-octet mass splittings. Subsequently, using Eq.~\ref{Eq:sigmapiN-sigma0-y}, one can obtain the strangeness content of the nucleon from a given experimental determination of $\sigma_{\pi N}$.

At LO, $\sigma_0=\hat{m}/(m_s-\hat{m})\left(M_\Xi+M_\Sigma-2M_N\right)\simeq27$ MeV. The NLO or $\mathcal{O}(p^3)$ corrections were first calculated by Gasser in 1982 using an early version of B$\chi$PT regularized with phenomenological form factors, giving $\sigma_0=35(5)$ MeV. This result was later bolstered by an $\mathcal{O}(p^4)$ HB calculation~\cite{Borasoy:1996bx} which obtained $\sigma_0=36(7)$. However, at $\mathcal{O}(p^4)$ several new unknown LECs contribute and they had to be determined in this calculation using model estimates, or even the value $\sigma_{\pi N}=45$ MeV as input. Besides, it is known that the HB approach suffers from a problematic convergence in $SU(3)$-flavor applications~\cite{Geng:2008mf} (for a recent review see~\cite{Geng:2013xn}). Nevertheless, and despite the possible problems in these numerical determinations, they have settled in the field, becoming an important source of distrust in ``relatively large'' values of $\sigma_{\pi N}$. Indeed, taking the Gasser result on $\sigma_0$ with the value of $\sigma_{\pi N}=64(8)$ MeV extracted from $\pi N$ scattering data by the GW group, one obtains a strangeness contribution to $M_N$ of about 300 MeV, a scenario that is considered implausible.

\begin{table*}[h]

\centering
\caption{Values of $\sigma_0$ for different B$\chi$PT approaches. \label{Table:sigma0}}
\begin{tabular}{c|c|cc|cc|}
\cline{2-6}
& &\multicolumn{2}{|c|}{ \raisebox{-1ex}[0.pt]{Octet $\mathcal{O}(p^3)$}}&\multicolumn{2}{|c|}{\raisebox{-1ex}[0.pt]{Octet+Decuplet $\mathcal{O}(p^3)$}} \\
&  \raisebox{2ex}[10.pt]{Tree level $\mathcal{O}(p^2)$}& HB & Covariant& HB-SSE& Covariant\\ 
\hline
\multicolumn{1}{|c|}{$\sigma_0$ [MeV]}&27&58(23)&46(8)&89(23)&58(8)\\
\hline
\end{tabular}
\end{table*}

These calculations were recently revisited in the context of B$\chi$PT framed in a covariant framework within the EOMS scheme and considering explicitly the contributions of the decuplet~\cite{Alarcon:2012nr}, which were neglected in previous works. The results of this analysis are summarized in Table~\ref{Table:sigma0}, where we present those corresponding to the EOMS and HB approaches, with and without decuplet contributions. The errors are obtained by the explicit computation of $\mathcal{O}(p^4)$ diagrams stemming from the $SU(3)$-breaking of the baryon masses in the $\mathcal{O}(p^3)$ diagrams. As we can see, the corrections to the LO result on $\sigma_0$ studied are large. The HB expansion has severe problems of convergence in the description of the sigma terms at ${\cal O}(p^3)$. Focusing in the following on the covariant results, we see that considering only the octet contributions the result is reasonably close to the classical result of Gasser~\cite{Gasser:1980sb}, whereas the new contributions given by the decuplet baryons are not negligible producing a $\sim$10 MeV rise on $\sigma_0$. In summary, this implies that a pion-nucleon sigma term of $\sim 60$ MeV is not at odds with a small strangeness content in the nucleon. In fact, plugging the result for $\sigma_{\pi N}$ from $\pi N$-scattering reported in the previous section, we obtain
\begin{equation}
y=0.02(13).
\end{equation} 

\section{Determinations from lattice QCD}

A theoretical determination of the sigma terms is accessible through the LQCD simulations. There are two possible strategies. The most straightforward one consists of directly evaluating the matrix elements~(\ref{Eq:STsDefinition}) in the lattice. However this is computationally very expensive due to the evaluation of the contributions of the current coupled to disconnected quark lines, which are expected to play an important role in the numerical determinations. The second and most widely used strategy is based on the Hellmann-Feynman theorem which relates the sigma-terms to the quark-mass derivatives of the nucleon mass.
Therefore, one can obtain $\sigma_{\pi N}$ and $\sigma_s$ by interpolating the physical nucleon mass with determinations of $M_N$ at different values of quark masses. One needs enough accurate determinations close to the physical point and the main difficulty lies in assessing the systematic effects originating from a specific choice of interpolators. In this sense, it is natural to use B$\chi$PT to perform these studies. Interpolators based on $SU(2)$ HB$\chi$PT up to $\mathcal{O}(p^3)$ and $\mathcal{O}(p^4)$ have become standard in the extrapolations of $M_N$ and determinations of $\sigma_{\pi N}$ performed by the lattice collaborations. 

Two main difficulties have been encountered in the development of this program based on B$\chi$PT. First, the extension of the formalism into a $SU(3)$ setup, describing the quark-mass dependence of the masses of all the octet (and decuplet) baryons and giving access to $\sigma_s$, has been troubled by the problematic convergence of the HB approach in this sector of the theory. Only after the application of approaches with cut-off regularization prescriptions~\cite{Young:2009zb} or in the covariant formalisms ~\cite{MartinCamalich:2010fp,Geng:2011wq,Semke:2011ez,Semke:2012gs,Ren:2012aj,Ren:2013dzt}, it has been possible to perform reliable $SU(3)_F$-B$\chi$PT extrapolations. Second, the systematic effects given by the decuplet degrees of freedom in the extrapolation of the baryon masses and on the value of the sigma-terms remains unclear. While the effect of $\Delta$ pieces on $\sigma_{\pi N}$ at $\mathcal{O}(p^3)$ in a $SU(2)$ calculation has been claimed to be negligible~\cite{Procura:2006bj}, a more thorough calculation of these effects up to $\mathcal{O}(p^4)$ contradicted this statement~\cite{WalkerLoud:2008bp}.

\begin{figure}[t]
\centering
\includegraphics[width=7cm]{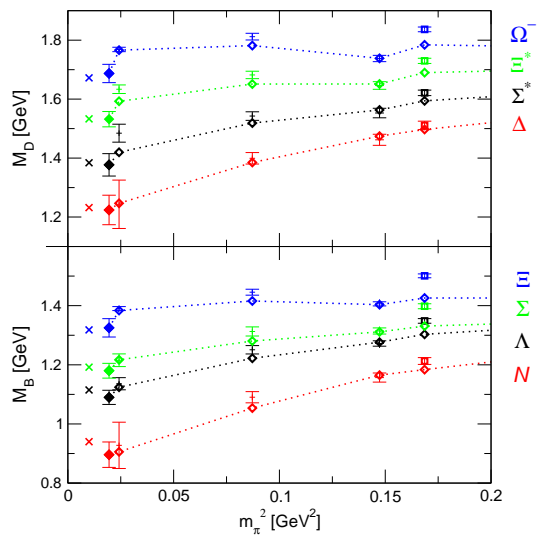}
\caption{Extrapolation of the PACS-CS results~\cite{Aoki:2008sm} on the lowest-lying baryon masses within the 
covariant formulation of $SU(3)_F$-B$\chi$PT up to $\mathcal{O}(p^3)$~\cite{MartinCamalich:2010fp}.\label{fig_extrapolationp3}}
\end{figure}

In order to address these two issues, we report on the results of the extrapolation of the octet (and decuplet) baryon masses obtained in $SU(3)$-B$\chi$PT in the EOMS scheme. In contrast with the results obtained using the HB expansion, it has been found that a good description of the LQCD results can be achieved within the Lorentz covariant approach to $SU(3)$-B$\chi$PT up to $\mathcal{O}(p^4)$ and considering the explicit inclusion of decuplet degrees of freedom~\cite{MartinCamalich:2010fp,Geng:2011wq,Ren:2012aj,Ren:2013dzt}. Moreover, an order-by-order improvement in the description of the lattice results on the octet baryon masses was found~\cite{MartinCamalich:2010fp,Ren:2013dzt}. Similar efforts in self-consistent formalisms up this accuracy have been reported by Semke and Lutz in~\cite{Semke:2011ez,Semke:2012gs} and also by the latter author in this conference.   

In Fig.~\ref{fig_extrapolationp3}, we show the quark mass dependence and extrapolation of the lowest-lying baryon masses in Lorentz covariant B$\chi$PT up to $\mathcal{O}(p^3)$ for the case of the analysis of the PACS-CS results~\cite{Aoki:2008sm}. The strategy followed was to fit the 4 (3) LECs appearing at this order for the octet (decuplet) baryons using the results of different LQCD collaborations. As it can be seen from the figure, the lattice results are well reproduced and the extrapolation to the physical point agrees with the experimental values within errors~\footnote{Notice that in these fits the experimental baryon masses are not included in the fit, so the results obtained at the physical point are a prediction of the extrapolation.}. The improvement obtained at this order in covariant $SU(3)$-B$\chi$PT in comparison with the description provided by the Gell-Mann-Okubo approach at $\mathcal{O}(p^2)$, stresses the relevance of the leading chiral non-analytical terms in the understanding of the nucleon mass from quark masses as light as those reached by PACS-CS~\cite{Aoki:2008sm} ($M_\pi\simeq156$ MeV), whereas the comparison with HB~\cite{MartinCamalich:2010fp} highlights the fact that the relativistic corrections greatly improve the description of the LQCD results on the baryon masses at heavier quark-masses.

\begin{table}[t]
\renewcommand{\arraystretch}{1.3}     
\setlength{\tabcolsep}{0.3cm}
\centering
\caption{Predictions on the $\sigma_{\pi N}$ and $\sigma_{sN}$ terms (in MeV) of the baryon-octet in covariant $SU(3)_F$-B$\chi$PT by fitting the LECs to the PACS-CS~\cite{Aoki:2008sm} or LHPC~\cite{WalkerLoud:2008bp} results. The errors are only statistical. \label{Table:ResSigmasB}}
\begin{tabular}{c|cc|cc|}
\cline{2-5}
&\multicolumn{2}{|c|}{PACS-CS}&\multicolumn{2}{|c|}{LHPC}\\
\cline{2-5}
\multicolumn{1}{c|}{}&No Dec.&Dec.&No Dec.&Dec.\\
\hline 
\multicolumn{1}{|c|}{$\sigma_{\pi N}$}&46(2)&59(2)&43(2)&61(2)\\
\multicolumn{1}{|c|}{$\sigma_{sN}$}&28(23)&$-7$(23)&6(20)&$-4$(20)\\
\hline
\end{tabular}
\end{table}

As a result of the determinations of the LECs from the fits, one can predict the nucleon sigma terms. In Table ~\ref{Table:ResSigmasB} we present the results on $\sigma_{\pi N}$ and $\sigma_{sN}$ after fitting the LECs to the PACS-CS and LHPC results. We also present the results that are obtained in fits with (Dec.) and without (No Dec.) the inclusion of decuplet resonances to discuss the systematic effects stemming from the treatment of these contributions. As we can see from this table, the results on $\sigma_{\pi N}$ in either case are very consistent with the analysis of the experimental data described in the previous section in the case of approximate fulfillment of the Zweig rule. This confirms, in a highly non-trivial fashion, the conclusions at $\mathcal{O}(p^3)$ in $SU(3)$-B$\chi$PT in the EOMS scheme derived exclusively from experimental data. In particular, it confirms that a scenario with a $\sigma_{\pi N}\simeq 60$ MeV can not be ruled out on the grounds of a small strangeness content of the nucleon at this order and that an irreducible uncertainty of about $15$ MeV originates from the omission of the decuplet.

\begin{figure}[t]
\centering
\includegraphics[width=14cm]{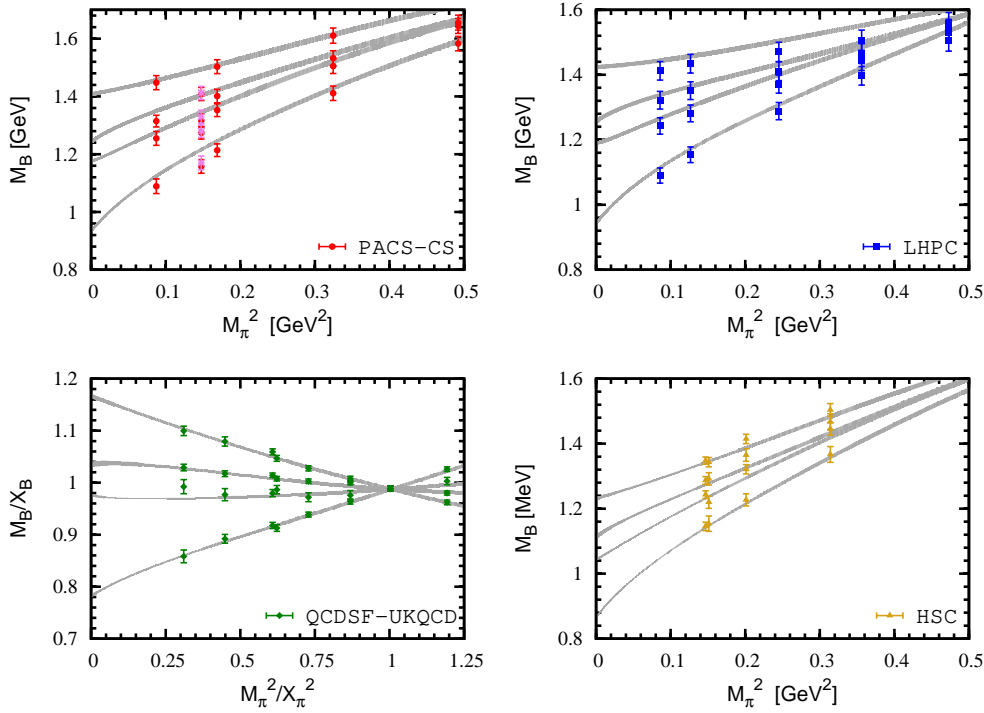}
\caption{Extrapolation of the PACS-CS results~\cite{Aoki:2008sm} on the lowest-lying baryon masses within the 
covariant formulation of $SU(3)_F$-B$\chi$PT up to $\mathcal{O}(p^4)$ and without decuplet degrees of freedom~\cite{Ren:2012aj}. \label{fig_extrapolationp4}}
\end{figure}

In order to settle the question of the strangeness $\sigma_s$  it is clear that one needs calculations at $\mathcal{O}(p^4)$. However, at this order a staggering  amount of 15 new unknown LECs enter the calculation and determining their values in a statistically sound fashion becomes a challenge. In fact, it seems impossible to constraint their values resorting to experimental data only and results from LQCD calculations have to be massively used. Nevertheless, first steps in this direction have been given and stable fits to global LQCD results on the baryon masses have been obtained. In particular, in the works by Semke and Lutz~\cite{Semke:2011ez,Semke:2012gs}, reported also in this conference, the usual chiral expansion is supplemented with another one in $1/N_c$ which allows to uncover hierarchies among the LECs and to reduce their total number. More general fits taking into account all the 19 LECs and also finite volume corrections have been presented in the EOMS scheme without~\cite{Ren:2012aj} and with the decuplet degrees of freedom~\cite{Ren:2013dzt} and using a total of 11 configurations at different quark masses and volumes (each of which contains four points for the $N$, $\Lambda$, $\Sigma$ and the $\Xi$). The resulting good description of the quark mass dependence of the lowest lying octet baryons in this approach is illustrated in Fig.~\ref{fig_extrapolationp4} where the results of these fits is plotted against PACS-CS~\cite{Aoki:2008sm}, LHPC~\cite{WalkerLoud:2008bp},  HSC~\cite{Lin:2008pr} and UKQCD~\cite{Bietenholz:2011qq} results. The NPLQCD~\cite{Beane:2011pc} results are also used but not plotted and the BMW results~\cite{Durr:2011mp} are not included in the analysis. As for the sigma terms, the situation is at the moment unclear. In this calculation the values $\sigma_{\pi N}=43(1)(6)$ MeV and $\sigma_{s N}=126(24)(54)$ MeV are reported~\cite{Ren:2012aj}, whilst in the calculation at the same order by Semke {\it et al.} the values $\sigma_{\pi N}=32(2)$ MeV and $\sigma_{s N}=22(20)$ MeV are given~\cite{Semke:2012gs} (see M.~Lutz's talk in this conference). Therefore, further efforts are required to understand these inconsistencies and to assess the convergence of the chiral expansion of these observables. Agreement with the results obtained with cut-off regularization~\cite{Young:2009zb} in the context of dimensional regularized approaches shall also be pursued. 

\section{Conclusions}

We have reviewed the status and potential of the modern approaches to B$\chi$PT by showing different recent determinations of the sigma terms. Besides being very important properties of the nucleon, they can be determined from different perspectives, based on $SU(2)$ or $SU(3)$ approaches along one direction but also using either experimental data or LQCD results along an orthogonal one. We have seen how the situation in our understanding of the $\pi N$ scattering data in a chiral framework has been greatly improved thanks to the application of modern Lorentz covariant techniques and dealing rigorously with the contributions of the $\Delta(1232)$. Although at the moment the resulting phenomenology still depends on the PW used as input, this progress offers the possibility of extracting the $\pi N$ scattering observables, and in particular $\sigma_{\pi N}$, in B$\chi$PT using directly the scattering cross-section data. As for the LQCD determinations, there has been much progress both in the quality of the LQCD results as well as on the accuracy of the B$\chi$PT calculations. Nevertheless, further work is needed to settle this issue from $\chi$PT perspective. On one hand, it would be desirable to revisit the lattice world data on $M_N$ using a $SU(2)$ framework to determine $\sigma_{\pi N}$. On the other, more data and, ideally, calculations and extrapolations on other observables in $SU(3)$ seem necessary to understand better the strangeness content of the nucleon at $\mathcal{O}(p^4)$.

\section{Acknowledgments}

I would like to thank the organizers for inviting me to this extremely interesting meeting. Also I would like to thank my collaborators J. M. Alarc\'on, L. S. Geng, J. Meng, J. A. Oller, X. L. Ren, H. Toki and M. J. Vicente Vacas who have contributed to the work presented in this talk. This work is funded by the Science Technology and Facilities Council (STFC) under grant ST/J000477/1, the Spanish Government and FEDER funds under contract FIS2011-28853-C02-01 and the grants FPA2010-17806 and Fundaci\'on S\'eneca 11871/PI/09.

\end{document}